\documentclass[twocolumn,showpacs,amsmath,amssymb,aps, prl,superscriptaddress]{revtex4-1}

\pdfoutput=1
\usepackage[]{graphicx}
\usepackage[breaklinks,colorlinks=true,citecolor=blue]{hyperref}
\usepackage[]{verbatim}
\usepackage[]{color}
\usepackage{overpic}
\usepackage{rotating}
\usepackage{mathtools}
\usepackage{appendix}
\usepackage{underscore}
\usepackage{times}
\usepackage{longtable}
\usepackage{multirow}
\usepackage{tabularx}
\usepackage[normalem]{ulem}

\makeatletter
\newsavebox{\@brx}

\newcommand{\llangle}[1][]{\savebox{\@brx}{\(\m@th{#1\langle}\)}%
  \mathopen{\copy\@brx\mkern2mu\kern-0.8\wd\@brx\usebox{\@brx}}}
\newcommand{\rrangle}[1][]{\savebox{\@brx}{\(\m@th{#1\rangle}\)}%
  \mathclose{\copy\@brx\mkern2mu\kern-0.8\wd\@brx\usebox{\@brx}}}

  \newcommand{\lllangle}[1][]{\savebox{\@brx}{\(\m@th{#1\langle}\)}%
  \mathopen{\copy\@brx\copy\@brx\mkern4mu\kern-0.7\wd\@brx\usebox{\@brx}}}
\newcommand{\rrrangle}[1][]{\savebox{\@brx}{\(\m@th{#1\rangle}\)}%
  \mathclose{\copy\@brx\copy\@brx\mkern4mu\kern-0.7\wd\@brx\usebox{\@brx}}}

\graphicspath{{./}{./}}

\begin{document}
\newcommand\redsout{\bgroup\markoverwith{\textcolor{red}{\rule[0.5ex]{2pt}{2.0pt}}}\ULon}
\title{Topological edge states in single layers of honeycomb materials with strong spin-orbit coupling}

\author{Andrei Catuneanu}
\affiliation{Department of Physics and Center for Quantum Materials, 
University of Toronto, 60 St. George St., Toronto, Ontario, M5S 1A7, Canada}

\author{Heung-Sik Kim}
\affiliation{Department of Physics and Center for Quantum Materials, 
University of Toronto, 60 St. George St., Toronto, Ontario, M5S 1A7, Canada}

\author{Oguzhan Can}
\affiliation{Department of Physics and Center for Quantum Materials, 
University of Toronto, 60 St. George St., Toronto, Ontario, M5S 1A7, Canada}

\author{Hae-Young Kee}
\email{hykee@physics.utoronto.ca}
\affiliation{Department of Physics and Center for Quantum Materials, 
University of Toronto, 60 St. George St., Toronto, Ontario, M5S 1A7, Canada}
\affiliation{Canadian Institute for Advanced Research / Quantum Materials Program,
Toronto, Ontario MSG 1Z8, Canada}

\begin{abstract}
  We study possible edge states in single layers of honeycomb materials such as  $\alpha$-RuCl$_3$ and A$_2$IrO$_3$ (A=Li, Na) with strong spin-orbit coupling (SOC).
These two dimensional systems exhibit linearly dispersing one-dimensional (1D) edge states when their 1D boundary forms a zig-zag shape. 
  Using an effective tight-binding model based on first principles band structure calculations including Hubbard U and SOC, we find degenerate edge states at the zone center and 
zone boundary. The roles of chiral symmetry and time-reversal symmetry are presented. 
  The implications to experimental signatures and the effects of disorder are also discussed.
\end{abstract}
\maketitle

{\it Introduction ---}
The honeycomb lattice has provided an excellent playground to explore novel physics in both weakly and strongly correlated systems.
Various phases such as the Dirac semimetal\cite{Castro2009}, Chern insulator\cite{Haldane1988}, and quantum spin Hall (QSH) insulator\cite{KaneMeleA,KaneMeleB} have been proposed in honeycomb lattices with and without spin-orbit coupling (SOC).
Moreover, strongly correlated honeycomb systems such as A$_2$IrO$_3$ (A = Na or Li)\cite{Singh2012}, in which the multi-orbital $t_{2g}$ model is mapped into a single orbital with pseudospin $j_\text{eff}=1/2$ owing to strong SOC\cite{BJKim2008,BJKim2009}, could in principle realize the exactly solvable Kitaev spin model\cite{Kitaev2006,Jackeli2009,WCKB2013,Chun2015,RLK2016}. 
Along this vein of thinking, $\alpha$-RuCl$_3$ (RuCl$_3$) was recently proposed as another promising candidate to exhibit Kitaev physics\cite{Plumb2014,KVCK2015}.
The unconventional magnetic ordering reported in RuCl$_3$, Na$_2$IrO$_3$, and Li$_2$IrO$_3$\cite{Liu2011,Ye2012,Choi2012,Singh2012,Chun2015,Jennifer2015,Kubota2015,Johnson2015,Cao2016} can be understood as a competition between Kitaev and other exchange terms\cite{Kimchi2011,Rau2014,Reuther2014} and thus the honeycomb lattice endowed with strong SOC has been one of hot topics in the community of spin liquid and topological phases. 

From the topological point of view, it was suggested that honeycomb Na$_2$IrO$_3$ can support the quantum spin Hall (QSH) state above the magnetic ordering temperature via complex second nearest neighbor hopping terms due to a combination of strong SOC and the edge sharing nature of the oxygen octahedra\cite{Shitade2009}, which leads to a gap
 at the $K$- and $K'$-point\cite{Shitade2009}.
This idea is an extension of the QSH in graphene\cite{KaneMeleA}, but it should be easier to observe in iridates because the SOC is much stronger.
Interestingly, an angle-resolved photoemission spectroscopy (ARPES) measurement on Na$_2$IrO$_3$ at room temperature reported linearly dispersing bands, but at the $\overline{\Gamma}$-point\cite{Alidoust2014}.
This observation remains unsolved, since a theoretical study based on {\it ab-initio} calculation found a trivial topological $\mathbb{Z}_2$ invariant for Na$_2$IrO$_3$\cite{Choong2012}, and there is currently no microscopic theory that supports such a surface state in the compound. Furthermore, other ARPES experiments \cite{Comin2012,Zhou2016} did not find any linearly dispering bands in Na$_2$IrO$_3$, perhaps indicating that differences in sample surfaces could be giving contrary results. 

\begin{figure}
  \centering
  \includegraphics[width=0.48\textwidth]{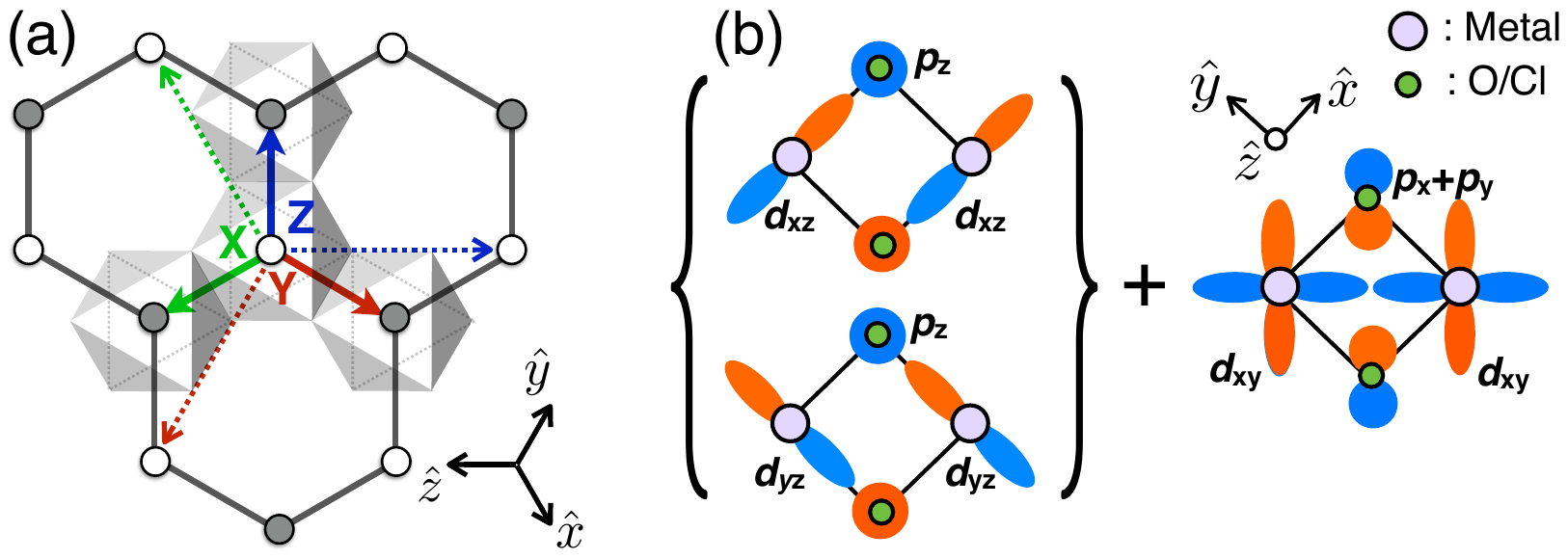}
  \caption{(Color online) (a) Schematics of a transition metal honeycomb layer, in which 
  edge-sharing MX$_6$ octahedra and the bonding between the NN and 2NN M sites are
  shown. Three inequivalent bonds (X, Y, Z) for the NN and 2NN are depicted in different colors
  (green, red, and blue, respectively). (b) NN $t_1$ term for the minimal model decomposed into
  two inequivalent $t_{\rm 2g}$-orbital hopping channels, in which participating $t_{\rm 2g}$- 
  and $p$-orbitals are shown. Note that the two channels cancel
 each other and yield a small $t_1$.}
  \label{fig:str}
\end{figure}

Here we study edge states of  two-dimensional single layers of $\alpha$-RuCl$_3$ and A$_2$IrO$_3$ 
by employing {\it ab-initio} electronic structure calculations including SOC and the Hubbard interaction U of 2-3eV, 
and analyzing an effective tight-binding (TB) model based on $j_\text{eff}=1/2$ pseudospins.
While the Hubbard interaction of 2-3eV favors a magnetic ground state, we impose a paramagnetic condition to study the topological nature of these materials in the absence of magnetic order,
which is thus applicable to the region above the magnetic ordering temperature. Furthermore, the Hubbard U enhances the $j_\text{eff}=1/2$ picture and makes the band gap bigger than without U\cite{KVCK2015}.
We estimate $j_\text{eff}=1/2$ TB parameters from our {\it ab-initio} calculations and find that the third neighbor $j_\text{eff}=1/2$ hopping channel ($t_3$) in these materials dominates over the first ($t_1$) and second ($t_2$) neighbor hopping channels because of cancellations of orbital overlaps within the $t_{2g}$ manifold in the first neighbor hopping channel.
Using these estimated TB parameters, we compute edge spectra in zig-zag, armchair and Klein edge configurations. We find that these compounds have topological edge states with linear dispersion at momentum points $k_x=0$ and $\pi$ points in the one-dimensional (1D) Brillouin zone only with the zig-zag boundary conditions provided that (1)  $t_3 > \frac{1}{3}t_1$ and (2) time-reversal symmetry (TRS) is preserved.

  \begin{figure*}[ht!]
  \centering
  \includegraphics[width=1.0\textwidth]{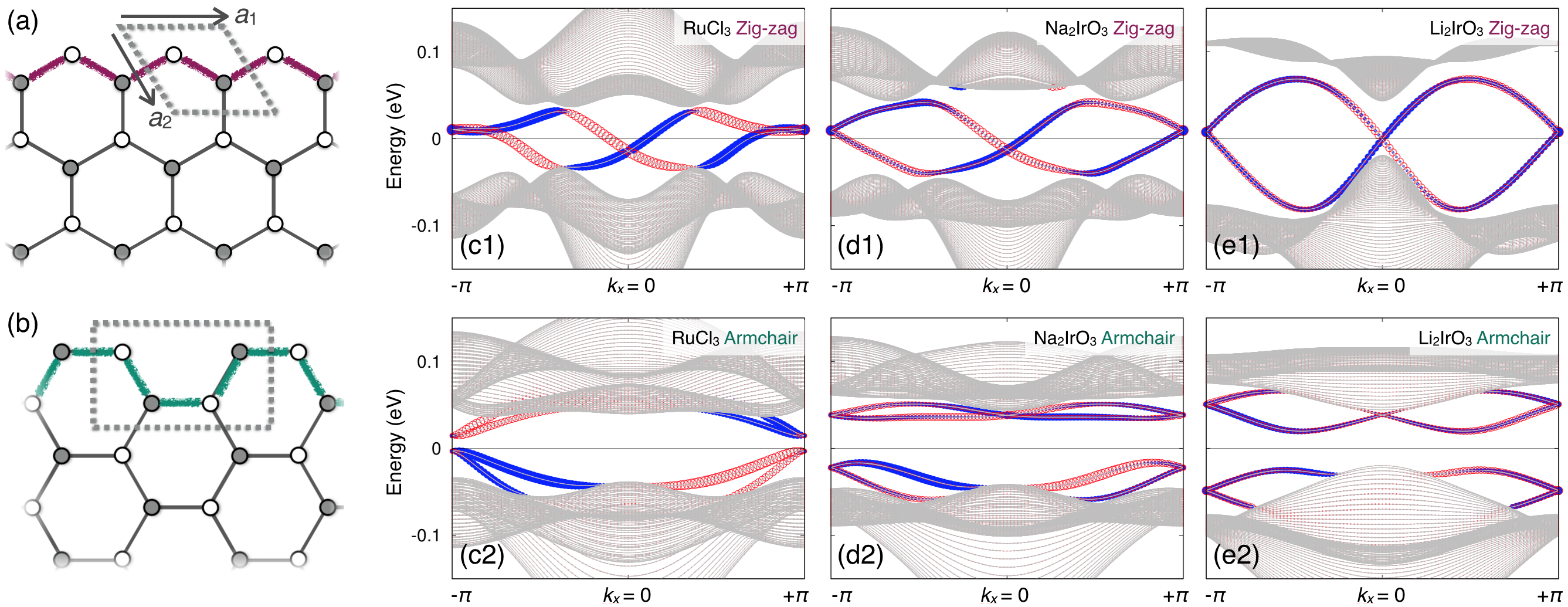}
  \caption{(Color online) (a,b) Depiction of (a) zig-zag and (b) armchair edges on the honeycomb lattice, 
  represented as violet and teal lines respectively. 
  The choice of the bulk unit cell for each edge configuration is shown as grey dashed lines. 
  (b-d) Zig-zag and armchair edge spectrum for (b) RuCl$_3$, (c) Na$_2$IrO$_3$, and (d) Li$_2$IrO$_3$. Upper (b1, c1, d1) and lower (b2, c2, d2) panels show zig-zag and armchair 
  bands, respectively. Underlaid blue/red colored symbols represent the pseudospin up/down edge weight of the state.
  }
  \label{fig:band}
\end{figure*}

{\it First principles band structure results and edge spectra ---}
We first perform local density approximation (LDA) calculations with SOC and Hubbard U to analyze the electronic band structure of RuCl$_3$ and A$_2$IrO$_3$\footnote{For the {\it ab-initio} density functional theory calculations we employed OpenMX code\cite{openmx}. See Supplementary Materials for further details of the calculations.}.
These compounds have crystal structures with $C2/m$ space group symmetry, consisting of weakly coupled  RuCl$_3$ and A$_{1/2}$IrO$_3$ layers stacked on top of each other \cite{Choi2012,Ye2012,Gretarsson2013,Johnson2015,Cao2016}.
Fig. \ref{fig:str}(a) is a schematic of each layer, in which edge-sharing metal (M) - anion (X) octahedra MX$_6$ form a honeycomb lattice of M ions.
Because of small monoclinic distortion, the threefold rotation symmetry is slightly broken and the first-nearest (NN) and second-nearest neighboring (2NN) Z-bonds in Fig. \ref{fig:str}(a) develop small differences compared to the X- and Y-bonds. 

\begin{table}
  \centering
  \setlength\extrarowheight{2pt}
  \newcolumntype{R}{>{\raggedleft\arraybackslash\hsize=.6\hsize}X}
  \newcolumntype{W}{>{\raggedleft\arraybackslash\hsize=1.0\hsize}X}
   \begin{tabularx}{0.48\textwidth}{lRRRW}
   \hline\hline   
   & $t_1$ & $t_2$ & $t_3$ & ${\bf T}^Z$  \\[5pt] \hline
   RuCl$_3$           & -4.3      & -8.0  & -44.4   &  (8.8, 8.6, 8.3)  \\   
   Na$_2$IrO$_3$ & -6.3      & -5.3 & -38.2    & (11.4, 11.4, -2.5) \\
   Li$_2$IrO$_3$   & -14.8    & -7.2 & -31.3   &  (20.8, 20.8, 0.8) \\  
    \hline\hline 
  \end{tabularx}
  \caption{(Units in meV)
  Hopping integrals $t_1$, $t_2$ and $t_3$ in Eq. \ref{eq:hops}, tabulated for three different compounds.
  Small bond dependence for hopping integrals is taken into account by averaging over the three inequivalent 
  bonds. For 2NN, only ${\bf T}^Z$ is shown and ${\bf T}^{X,Y}$ are presented in the Supplementary Materials.}
  \label{tab:hops}
\end{table}

Previous theoretical studies on the above compounds showed that, due to SOC, electronic structures
near the Fermi level can be mapped to a minimal single-orbital model of the $j_{\rm eff}=1/2$ states 
on the M-sites\cite{Shitade2009,HSK2013,KVCK2015}, which can be expressed as
\begin{align}
  H = &\Big(
  	t_1\sum_{\langle i,j\rangle}C^\dagger_{i}C_{j} 
       +t_2\sum_{\langle\langle i,j\rangle\rangle}C^\dagger_{i}C_{j} 
       +t_3\sum_{\langle\langle\langle i,j\rangle\rangle\rangle} C^\dagger_i C_j\nonumber \\
  &+ i\sum_{\langle\langle i,j\rangle\rangle\in\gamma}\nu_{ij}C^\dagger_i {\bf T}^{\gamma} \cdot \boldsymbol{\sigma}C_j
  	\Big) +h.c.
  \label{eq:hops}
\end{align}
where $C_i$ is the annihilation operator for the $j_{\rm eff}=1/2$ pseudospin spinor 
$(c_{i\uparrow}, c_{i\downarrow})$ at site $i$, $\gamma$ denotes the bond type as in Fig. \ref{fig:str}(a)  
and ${\bf T}^\gamma \equiv (t^\gamma_{2x}, t^\gamma_{2y}, t^\gamma_{2z})$ are bond-dependent
2NN hopping terms with $\boldsymbol{\sigma}\equiv(\sigma_x, \sigma_y, \sigma_z)$ being Pauli matrices 
for the pseudospin subspace. 
The phases $\nu_{ij} = \pm1$ depend on A ($+1$) or B ($-1$) sublattice hopping.
Projecting the $j_{\rm eff}=1/2$-dominated subspace onto the minimal model using
Wannier orbitals for each compound in our {\it ab-initio} calculations yields the model 
parameters tabulated in Table \ref{tab:hops}. A common feature is the predominating
third-nearest neighbor (3NN) $t_3$ and 2NN ${\bf T}^\gamma$ terms compared to the NN $t_1$ and 2NN $t_2$. 
The small value of $t_1$, which seems 
counterintuitive considering the size of NN $t_{\rm 2g}$ hopping terms ($\sim$ 100 meV)
\cite{Foyevtsova2013,KVCK2015,Winter2016}, 
is the result of cancellation within the NN $t_{\rm 2g}$ hopping channels 
originating from the edge-sharing geometry and the form of the $j_{\rm eff}=1/2$ orbitals. 
Fig. \ref{fig:str}(b) shows the two most significant channels participating in $t_1$, 
the $\delta$- and $\sigma$-like overlap integrals between the neighboring $t_{\rm 2g}$
orbitals. Since the two channels have opposite signs to each other, cancellation between 
them yields vanishingly small values of $\sim$ 10 meV for $j_{\rm eff}=1/2$ $t_1$ in all three compounds. 
On the contrary, 2NN and 3NN channels are enhanced by the presence of 
different channels, in addition to the
$j_{\rm eff}=1/2$ -- $3/2$ -- $1/2$ virtual processes which are effective beyond the 
NN channels (See Supplementary Materials for further details). Note that the small $t_1$ 
leads to the weak NN Heisenberg term in the spin Hamiltonian as reported in Ref. \onlinecite{Kim2016,Winter2016}.

\begin{table}
 \centering
  \begin{tabular}{ l | c | c}
     & $\vec{d}(\textbf{k})$ & $\Gamma$\\ 
    \hline\hline
    $d_0(\textbf{k})$ & $2t_2(\cos k_x + \cos k_y + \cos (k_x-k_y))$ & $I_4$\\[0.1cm]
    $d_1(\textbf{k})$ & $t_3(\cos(k_y-2k_x) + 2\cos k_y)$ & $I_2\otimes\tau_x$ \\
    & $t_1(1+\cos(k_y-k_x)+\cos k_x)$ \\[0.1cm]
    $d_2(\textbf{k})$ & $-t_3\sin(k_y-2k_x)-t_1(\sin(k_x-k_y)+\sin k_x)$ & $I_2\otimes\tau_y$\\[0.1cm]
    $d_3(\textbf{k})$ & $-2(t_{2x}\sin k_x -t_{2z}\sin k_y + t_{2y}\sin(k_y-k_x))$ & $\sigma_x\otimes\tau_z$ \\[0.1cm]
    $d_4(\textbf{k})$ & $-2(t_{2y}\sin k_x -t_{2x}\sin k_y + t_{2z}\sin(k_y-k_x))$ & $\sigma_y\otimes\tau_z$ \\[0.1cm]
    $d_5(\textbf{k})$ & $-2(t_{2z}\sin k_x -t_{2y}\sin k_y + t_{2x}\sin(k_y-k_x))$ & $\sigma_z\otimes\tau_z$
  \end{tabular}
  \caption{Coefficients of the $\Gamma$ matrices in Eq. \eqref{eq:bloch-hamiltonian}. We have chosen a momentum space basis such that $\textbf{k} = k_x\textbf{b}_1+k_y\textbf{b}_2$, where $\textbf{b}_i$ are the reciprocal lattice vectors corresponding to the primitive lattice vectors ${\bf a}_{1,2}$ chosen as in Fig. \ref{fig:band}(a).}
\end{table}

Fig. \ref{fig:band} shows the edge spectra of RuCl$_3$, Na$_2$IrO$_3$, and Li$_2$IrO$_3$ for zig-zag and armchair boundary configurations. Linearly dispersing edge states exist on the zig-zag boundary (Fig. \ref{fig:band}(c-e1)) while the armchair boundary does not support any such states (Fig. \ref{fig:band}(c-e2)).
The degenerate points at $k_x = 0$ and $\pi$ on the zig-zag boundary are protected by TRS, which will be discussed in detail in the following section. 

\begin{figure}[!ht]
 \centering
 \includegraphics[width=0.49\textwidth]{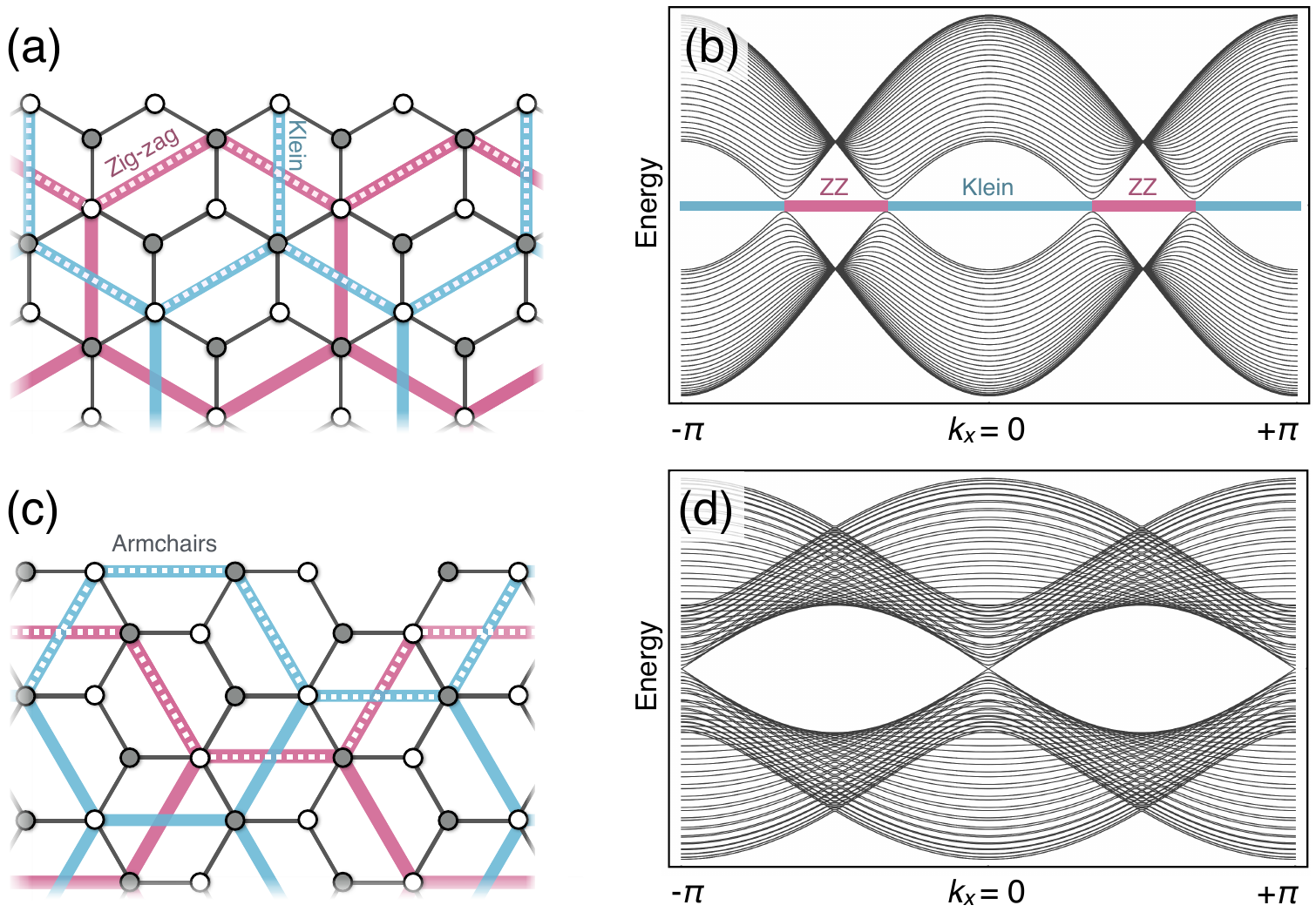}
 \caption{(Color online) Understanding the $t_3$-only limit for zig-zag and armchair boundary configurations. (a) Two big honeycombs form zig-zag (pink) and Klein (blue) boundary configurations on the zig-zag boundary of the underlying lattice. (b) Zig-zag boundary spectrum with colored edge states corresponding to the two big honeycombs. (c) The two big honeycombs both form armchair boundary configurations when the underlying lattice forms an armchair boundary. (d) No edge states are observed in the armchair spectrum.}
 \label{fig:3}
\end{figure}

{\it Tight-binding model and edge states ---}
To understand the nature of the Hamiltonian (\ref{eq:hops}) and the edge spectra, let us consider the TB Hamiltonian which takes the form $H = \sum_\textbf{k} \psi^\dagger_\textbf{k}h(\textbf{k})\psi_\textbf{k}$; 
\begin{equation}
  h(\textbf{k}) = d_0(\textbf{k})I_4 + \vec{d}(\textbf{k})\cdot\vec{\Gamma}.\label{eq:bloch-hamiltonian}
\end{equation}
We define the momentum space spinor in terms of $j_\text{eff}=1/2$ operators by $\psi_\textbf{k} = (c_{\textbf{k}A\uparrow}, c_{\textbf{k}B\uparrow}, c_{\textbf{k}A\downarrow}, c_{\textbf{k}B\downarrow})$. The components of the vector $\vec{d}(\textbf{k})$ are given in Table II and $\vec{\Gamma} = (I_2\otimes\tau_x,I_2\otimes\tau_y,\sigma_x\otimes\tau_z,\sigma_y\otimes\tau_z,\sigma_z\otimes\tau_z)$ is a vector of $4\times 4$ matrices composed of sublattice $\tau$ and pseudospin $\sigma$ Pauli matrices.
For the moment we ignore the identity term $d_0$ because it merely shifts the energy spectrum by a momentum-dependent constant and does not affect the existence of the edge states.

To comprehend the linearly dispersing edge states and their dependence on the boundary conditions, we provide a systematic analysis of the TB model below.  Recall that the NN model has been widely studied\cite{Fujita1996,Nakada1996,Ryu2002,Delplace2011,Takahashi2013} and it harbors flatband edge states on the zig-zag boundary for $|k_x| \geq 2\pi/3$, with no edge states appearing at $k_x = 0$. 
The current results go beyond the NN model. We note that the 3NN-only TB model involves two separated bigger honeycombs as shown in Fig. \ref{fig:3}(a,c) with red and blue colors, and that the edge states at $k_x=0$ may come from the Klein boundary of the blue honeycomb in Fig. \ref{fig:3}(a). To prove this rigorously, we first investigate the $t_3$-only limit, then we show the effects of adding the NN $t_1$ term, followed by spin-dependent 2NN terms.

The $t_3$-only limit given by $h(\textbf{k}) = d_1(\textbf{k})\tau_x + d_2(\textbf{k})\tau_y \equiv [d_1(\textbf{k}), d_2(\textbf{k}), 0]$ with $t_1 = 0$ posseses chiral symmetry because it anticommutes with $\tau_z$.
We follow the winding number prescription discussed in an earlier work\cite{Ryu2002} and project the Hamiltonian onto different edge configurations by cutting the lattice as in Fig. \ref{fig:3}(a,c) and Fourier transforming along the x direction to obtain a set of 1D Hamiltonians parameterized by $k_x$, denoted by $h_{k_x}(k_y)$. As $k_y$ is varied from 0 to $2\pi$, $h_{k_x}(k_y)$ traces a loop in $\tau$-space and if it contains the origin, the existence of degenerate edge states is guaranteed.
The $t_3$-only model on the zig-zag boundary gives $h_{k_x}(k_y) = -t_3[\cos(k_y-2k_x)+2\cos k_y,-\sin(k_y-2k_x),0]$, which always contains the origin thus giving flatbands for all $k_x$ in the 1D Brillouin zone, as evidenced in Fig. \ref{fig:3}(b).
For the armchair edge, $h_{k_x}(k_y) = -t_3[\cos(k_x) + 2\cos(k_x-2k_y), \sin k_x, 0]$ which does not enclose the origin for any $k_x$ so no edge states appear as shown in Fig. \ref{fig:3}(d). On the Klein edge, the two big honeycombs terminate as zig-zag edges so no edge states appear at $k_x = 0$. The edge spectrum of the $t_3$-only limit does not qualitatively change when adding $t_1$, except for small regions of $k_x$ so long as $t_3 > \frac{1}{3}t_1$. We confirm that the corresponding winding number for all $k_x$ where the edge state exists is $\pm 1$.

We find that by adding the $d_5(\textbf{k})\sigma_z\tau_z$ term, the system is in a QSH phase with spin Chern number 2. The edge spectrum on a zig-zag boundary is shown in Fig. \ref{fig:4}(a). When $t_1$ is increased so that $t_3 = \frac{1}{3}t_1$, a band inversion occurs in the bulk at the M point and the model enters a non-trivial $\mathbb{Z}_2$ phase. 
The condition that $t_3 > \frac{1}{3}t_1$ is satisfied in all three compounds due to the cancellation of orbitals at the NN level discussed in the previous section.
$h(\textbf{k})$ can be block-diagonalized into pseudospin up ($+$) and down ($-$) sectors.
Within each sector we project $h(\textbf{k})$ onto the zig-zag boundary to give $h_{k_x,\pm}(k_y) = [d_1(\textbf{k}), d_2(\textbf{k}), \pm d_5(\textbf{k})]$ and find that $h_{k_x,\pm}$ wind around the origin at only $k_x = 0, \pi$ and a third point whose position depends on the TB parameters.
We therefore expect zero-energy edge states for each pseudospin at these points in the 1D Brillouin zone. The zero-energy edge states, denoted by black dots in Fig. \ref{fig:4}(a), are two-fold degenerate Kramers pairs. 

\begin{figure}
  \includegraphics[width=0.5\textwidth]{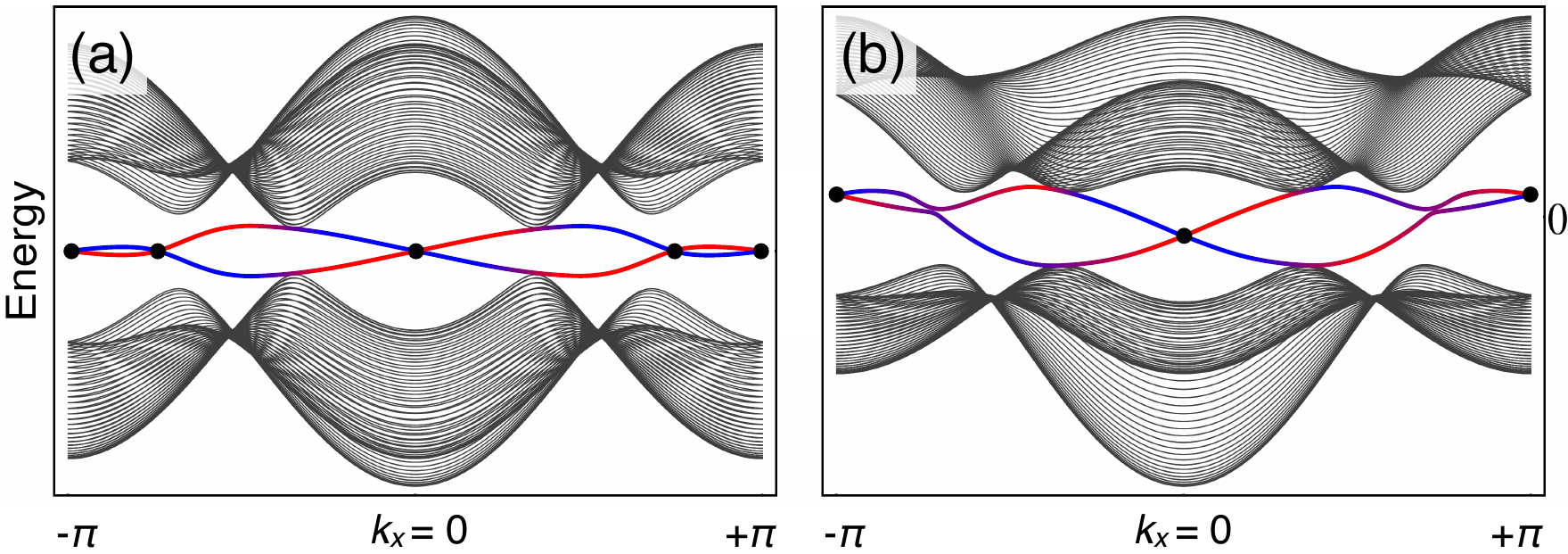}
  \caption{
  (Color online) Zig-zag edge spectra using {\it ab-initio} determined RuCl$_3$ TB parameters $t_1 = -4.3$ meV, $t_2 = -8.0$ meV, $t_3 = -44.4$ meV, $t_{2x} = 8.8$ meV,  $t_{2y} = 8.6$ meV and $t_{2z} = 8.3$ meV. We plot (a) $t_1+t_3$ model adding $\sigma_z\tau_z$ term and (b) all terms in $H$. Dark circles indicate the location of the degenerate edge states and blue/red coloring denotes up/down pseudospin.}
  \label{fig:4}
\end{figure}

When including the spin-flip terms $d_3(\textbf{k})\sigma_x\tau_z$ and $d_4(\textbf{k})\sigma_y\tau_z$, only the degenerate edge states 
at time-reversal-invariant-momentum (TRIM) $k_x = 0$ and $\pi$ remain on the zig-zag boundary.
The edge spectrum of the Hamiltonian on the zig-zag boundary using {\it ab-initio} TB parameters for RuCl$_3$ is plotted in Fig. \ref{fig:4}(b). The corresponding 1D Hamiltonian is given by $h_{k_x}(k_y) = [d_1(\textbf{k}), d_2(\textbf{k}), d_3(\textbf{k})\sigma_x+d_4(\textbf{k})\sigma_y+d_5(\textbf{k})\sigma_z]$. We rotate the pseudospin basis so that $d_3(\textbf{k})\sigma_x+d_4(\textbf{k})\sigma_y+d_5(\textbf{k})\sigma_z \rightarrow \sqrt{d_3^2(\textbf{k}) + d_4^2(\textbf{k}) + d_5^2(\textbf{k})}\sigma_z \equiv \Delta(\textbf{k})\sigma_z$ and construct $h_{k_x}^\pm(k_y) = [d_1(\textbf{k}), d_2(\textbf{k}), \pm\Delta(\textbf{k})]$ as before. Using this basis, we note that at $k_x = 0$ and $\pi$ we have $\Delta(\textbf{k}) \propto \sin k_y$. At these momenta, the 1D Hamiltonians describe ellipses that contain the origin in $\tau$-space so that we expect a pair of degenerate zero-energy edge states at $k_x = 0$ and $\pi$. This is confirmed numerically in Fig. \ref{fig:4}(b), in which the identity term $d_0(\textbf{k})$ is included.
 Adding the identity term $d_0 ({\bf k})$ shifts the edge states at $k_x = 0$ below the Fermi energy and those at $k_x=\pi$ above the Fermi energy as shown in Fig. \ref{fig:4}(b) since the chiral symmetry is broken, but the degeneracy of the Kramers pairs is protected by TRS. 
Our TB model was constructed assuming a perfect $C_3$ symmetry; however, monoclinic distortion in Na$_2$IrO$_3$, Li$_2$IrO$_3$ and honeycomb $\alpha$-RuCl$_3$ slightly breaks the $C_3$ symmetry but it does not change our conclusion.

{\it Discussion and conclusion ---}
In this paper we have explored the edge spectra in single layers of the honeycomb $\alpha$-RuCl$_3$ and family of iridates A$_2$IrO$_3$ (A = Na, Li).
We find from {\it ab-initio} band structure  analysis that these compounds, with similar geometry, all possess linearly dispersing edge states at $k_x = 0$ and $\pi$ on zig-zag boundary configurations. 
The degeneracy at TRIM points is protected by TRS. We used an effective TB model based on $j_\text{eff}=1/2$ pseudospins to describe the band structure and discovered that 3NN hopping integrals are necessary to adequately describe the band structure near the Fermi level. In fact, the 3NN hopping is crucial and should satisfy the condition $t_3 > \frac{1}{3}t_1$ to have degenerate edge states at $k_x = 0$ on the zig-zag boundary. 
Furthermore, the overlap integrals in the $t_{2g}$ manifold are appreciably affected by lattice distortion, as has been thoroughly investigated in a recent {\it ab-initio} band structure study\cite{Kim2016}. We found that the condition $t_3 > \frac{1}{3}t_1$ is robust for the optimized lattice parameters in all three materials and the edge states should be present if TRS is present. The current study also implies that effective spin models for these materials with strong SOC requires a careful analysis of further neighbor exchange interactions and their dependence on lattice optimization. For example, a recent study found that the 3NN Heisenberg term is important in determining the magnetic ground state in Na$_2$IrO$_3$ and RuCl$_3$\cite{Winter2016}.

Since there are two Kramers pairs of edge states, the effects of disorder and interactions deserve some discussion.
The stability of the $\mathbb{Z}_2$ topological insulator is limited to single-particle scattering processes because the backscattering between Kramers partners is prevented by TRS\cite{KaneMeleA}.
This was pointed out by Xu and Moore in Ref. \cite{XuMoore2008}. There it was shown that multi-particle scattering processes may localize the edge modes in the case of an odd number
of Kramers pairs. On the other hand, for two Kramers pairs of edge states such as in the current case, it was found that the edge states are stable if repulsive interactions
between Kramers partners are stronger than interactions between Kramers pairs. We note that the two Kramers pairs in these materials are separated by large momentum, thus we expect that
the intrapair interactions are stronger than interpair interactions.
 
Our result frames a recent ARPES measurement on Na$_2$IrO$_3$ which reported linearly dispersing surface states at the $\overline{\Gamma}$ point\cite{Alidoust2014}. 
We speculate that the surface states centered at the $\overline{\Gamma}$ observed in Ref. \onlinecite{Alidoust2014} originate from the zig-zag boundary of Na$_{1/2}$IrO$_3$ single layers exposed on the surface of Na$_2$IrO$_3$ above the magnetic transition temperature. 
We propose the fabrication of single layers of these honeycomb systems to search for the first realization of 1D topological edge states in 2D honeycomb transition metal materials with strong SOC.
In particular, $\alpha$-RuCl$_3$ is the best candidate among them since the interlayer coupling is extremely weak, possibly allowing for the creation of single layers by the graphene scotch tape method.

{\it Acknowledgements ---}
This work was supported by the NSERC of
Canada and the center for Quantum Materials at the University of
Toronto.  Computations were mainly performed on the GPC supercomputer
at the SciNet HPC Consortium. SciNet is funded by: the Canada
Foundation for Innovation under the auspices of Compute Canada; the
Government of Ontario; Ontario Research Fund - Research Excellence;
and the University of Toronto.

\bibliography{rucl3-chiral}

\end{document}